# Probing spin-phonon interactions in silicon carbide with Gaussian acoustics

Samuel J. Whiteley[1,2]*, Gary Wolfowicz[1,6]*, Christopher P. Anderson[1,2], Alexandre Bourassa[1], He Ma[1,4], Meng Ye[1], Gerwin Koolstra[2,5], Kevin J. Satzinger[1,7], Martin V. Holt[8], F. Joseph Heremans[1,3], Andrew N. Cleland[1,3], David I. Schuster[2,5], Giulia Galli[1,3,4], David D. Awschalom[1,3]†

[1] Institute for Molecular Engineering, University of Chicago, Chicago, IL, 60637, USA

[2] Department of Physics, University of Chicago, Chicago, IL 60637, USA

[3] Institute for Molecular Engineering and Materials Science Division, Argonne National Laboratory, Argonne, IL 60439, USA

[4] Department of Chemistry, University of Chicago, Chicago, IL 60637, USA

[5] The James Franck Institute, University of Chicago, Chicago, IL 60637, USA

[6] WPI-Advanced Institute for Materials Research (WPI-AIMR), Tohoku University, Japan

[7] Department of Physics, University of California, Santa Barbara, Santa Barbara, CA 93106, USA

[8] Center for Nanoscale Materials, Argonne National Laboratory, Argonne, IL 60439, USA

* These authors contributed equally to this work

† Corresponding author: awsch@uchicago.edu

## Abstract

Hybrid spin-mechanical systems provide a platform for integrating quantum registers and transducers. Efficient creation and control of such systems require a comprehensive understanding of the individual spin and mechanical components as well as their mutual interactions. Point defects in silicon carbide (SiC) offer long-lived, optically addressable spin registers in a wafer-scale material with low acoustic losses, making them natural candidates for integration with high quality factor mechanical resonators. Here, we show Gaussian focusing of a surface acoustic wave in SiC, characterized by a novel stroboscopic X-ray diffraction imaging technique, which delivers direct, strain amplitude information at nanoscale spatial resolution. Using *ab initio* calculations, we provide a more complete picture of spin-strain coupling for various defects in SiC with $C_{3v}$ symmetry. This reveals the importance of shear for future device engineering and enhanced spin-mechanical coupling. We demonstrate all-optical detection of acoustic paramagnetic resonance without microwave magnetic fields, relevant to sensing applications. Finally, we show mechanically driven Autler-Townes splittings and magnetically forbidden Rabi oscillations. These results offer a basis for full strain control of three-level spin systems.

# Main text

Hybrid quantum systems[1] leverage the strengths of various modalities of representing quantum information, including optical photons for sending quantum states across long distances, spins for information storage, and microwave superconducting circuits for computation, with the potential of using nanomechanics as an intermediary quantum bus. For instance, coherently exchanging quantum information between optically-active defect spins and mechanical resonators[2] provides a route to couple optical photons to microwave frequency phonons in a hybrid quantum system. Optically-active defect spins in SiC, such as the neutral divacancy[3] (VV), have recently been shown to support long-lived spin states[4,5,6], a variety of quantum controls[7], and spin-photon interfaces[8] compatible with quantum entanglement protocols. Importantly, SiC is a piezoelectric material and supports mature fabrication processes for production of high quality micro-electromechanical systems (MEMS). Although progress has been made coupling spins to mechanics in similar defect systems, including the NV center in diamond with strain tuning[9,10] and mechanical driving[11,12,13], defects in SiC are well positioned to solve the materials challenges of coherently manipulating spins with strain and strongly coupling spins with phonons.

While static strain will generate shifts in ground state (S = 1) energy sublevels, resonant AC strain can coherently drive electron spin transitions. Large in-plane dynamic strains can be generated by surface acoustic wave (SAW) devices, which are well developed for radio frequency (RF) filters and offer simple engineering approaches for fabricating low loss resonators. SAW devices have also been proposed as hybrid quantum transducers[14] and used to demonstrate coupling to superconducting qubits[15,16,17] along with optomechanical interactions involving defect excited states[18,19].

Here, we demonstrate acoustically driven $\Delta m_s = \pm 1$ spin transitions, where $m_s = 0, \pm 1$ is the spin projection, on VV spin ensembles in 4H-SiC. We further demonstrate $\Delta m_s = \pm 2$ spin transitions through the Autler-Townes effect, mechanical Rabi oscillations, and comparing the relative coupling strengths of inequivalent VV defects. These results are well described by our theoretical model developed from a combination of direct experimental observations and Density Functional Theory (DFT) calculations of anisotropic spin-strain coupling coefficients. We find that uniaxial strain and shear drive VV spins with coupling strengths of similar magnitude, but with generally different relative phase and selection rules. These experiments utilize a patterned Gaussian SAW phonon resonator that focuses strain and reduces resonator mode volumes in analogy to Gaussian optics. To image the mechanical modes of our Gaussian SAW devices, we use a unique nanoscale scanning X-ray diffraction technique that directly measures RF lattice perturbations. In addition, spatial responses of Autler-Townes splittings are well explained by ensemble averaging shear and uniaxial strain from the SAW mode. Shear provides an important way of controlling three-level spins (qutrits) with phonons and opens avenues for coupling spins with MEMS.

We first describe device design and characterization with a nanoscale X-ray diffraction imaging method, followed by spin manipulation. To amplify the piezoelectric response of the SiC substrate, we use a thin, sputtered aluminum nitride (AlN) layer on the SiC surface before fabricating a SAW resonator to create RF mechanical strain. Standard planar SAW resonator designs span wide apertures, often greater than 100 acoustic wavelengths (λ), distributing the strain across large crystal areas. Since AlN and 4H-SiC have isotropic in-plane Rayleigh wave velocities[20] (5790 and 6830 m/s, respectively), we fabricate simple Gaussian geometries, inspired by Gaussian optics, to focus strain while also suppressing acoustic diffraction losses (Fig. 1a,b). A patterned aluminum interdigitated transducer (IDT) transmits SAWs (λ = 12 μm), while grooves in AlN form Bragg gratings that act as SAW cavity mirrors to support a resonator frequency $\omega_m/2\pi \approx$ 560 MHz and loaded quality factor of ~16,000 (Fig. 1c) at 30 K. In our experiments the Gaussian geometries for enhanced strain focusing and reduced resonator mode volumes facilitate larger strains for fast coherent manipulation of electron spin states.

To directly visualize the Gaussian mechanical mode, we use stroboscopic scanning X-ray diffraction microscopy[21] (s-SXDM) to image the phonons with nanoscale resolution. This technique utilizes coherent X-rays from a synchrotron radiation light source, generated at 8 keV and focused to a 25 nm spot size (FWHM), and Bragg diffraction contrast to enable local measurements of lattice curvature and strain along a particular crystal orientation. We frequency match a Gaussian IDT's RF excitation to the timing structure of the synchrotron storage ring in order to measure the peak-to-peak amplitude of the acoustic standing wave. Due to the frequency matching requirements for s-SXDM, we use an IDT without a cavity (see Methods for device specifications), which is designed to produce a spatial strain mode similar to resonators used in spin experiments. Scanning the nano-focused X-ray beam in real space clearly shows the SAW profile (Fig. 1d) is consistent with the fabricated geometry and approximately nanometer Rayleigh wave displacements. The dynamic transverse lattice slope (Fig. 1e), caused by a local lattice plane tilt towards the ±$y$ direction, is expected from a Gaussian focus and SAW confinement. These X-ray measurements confirm that the SAW out-of-plane displacement (in phase with the in-plane uniaxial strain required for spin driving) is maximized at the resonator's precise center.

Electron spin ground state sublevels of VV defects are typically measured using optically detected magnetic resonance (ODMR) with $\Delta m_s$ = ±1 transitions magnetically driven by microwave fields. Due to the defect's intersystem crossing, ODMR probes the spin projections of $|\pm 1\rangle$ versus $|0\rangle$ through changes in photoluminescence (PL). The ground state spin Hamiltonian neglecting hyperfine interactions takes the form,

$$H/h = \gamma B \cdot S + S \cdot D \cdot S$$

Where $h$ is the Planck constant, $\gamma$ is the electron gyromagnetic ratio (≈ 2.8 MHz/G), $B$ is the external magnetic field vector, and $D$ is the zero-field splitting tensor. In the absence of lattice strain, the VV spin-spin interaction simplifies to $D_0 S_z^2$ where $D_0$ ~ 1.336 and 1.305 GHz, depicted in Fig. 2a, for c-axis oriented defect configurations[22] *hh* and *kk*, respectively. The zero-field splitting is sensitive to local lattice perturbations[23] such as thermal disorder, an applied electric field, or strain. When the lattice is perturbed by a small strain, characterized by a tensor $\varepsilon_{kl}$, the zero-field splitting tensor is generally modified by $\Delta D_{ij} = G_{ijkl}\varepsilon_{kl}$ where $G_{ijkl}$ is the spin-strain coupling tensor. The symmetry of the spin-strain coupling tensor is determined by the local C$_{3v}$ symmetry of the *hh* and *kk* configurations for VV (Fig. S1) and also applies to the NV center in diamond[24]. We utilize off-diagonal Hamiltonian elements containing $\Delta D_{ij}$ to drive resonant spin transitions with phonons, and consider both $\Delta m_s = \pm 1$ and $\pm 2$ transitions for full ground state S = 1 spin control.

We first demonstrate mechanical driving of $\Delta m_s = \pm 1$ spin transitions with the Gaussian SAW resonator. The point group symmetries of the VV in SiC allow for non-zero spin-strain coupling coefficients for zero-field splitting terms that contain the anticommutators $\{S_x, S_z\}$ and $\{S_y, S_z\}$ (derivation in supplementary materials). In order to probe acoustic paramagnetic resonance, we tune the axial magnetic field (B$_0$) such that the spin $|0\rangle$ to $|-1\rangle$ transition frequency is matched with the SAW resonator (Fig. 2a). It is critical to design an experimental measurement sequence insensitive to stray magnetic fields from electrical currents in the IDT. To disentangle these effects, we use an interlaced RF pump/laser probe sequence as well as lock-in amplification to measure the difference in luminescence when the spin resonance frequency is shifted away from the cavity resonance via modulation of B$_0$ with a small coil. Spin rotations are primarily driven and detected during the SAW cavity ring down period without RF driving, although the spin ensemble will also encounter some residual magnetic resonance when the RF drive is turned back on due to lingering optical-spin polarization. We detect higher PL contrast when the RF drive is matched to our SAW cavity resonance (Fig. 2b), whereas smaller, residual PL contrast is detected when the RF drive is far off SAW resonance. When the PL contrast is normalized by ODMR experiments from magnetic driving, the *kk*:*hh* mechanical drive rate ratio is 0.89 ± 0.10, which agrees with our theoretical model and DFT calculations (ratio ~ 1.0) where shear couples more strongly to $\Delta m_s = \pm 1$ transitions than does uniaxial strain (Table S2). The transverse spatial dependence (Fig. 2c) confirms that the PL contrast we measure on resonance matches our Gaussian resonator's mechanical mode shape. Magnetic field driving from the IDT, on the other hand, results in a flat profile (two-dimensional spatial mapping is shown in Fig. S3). The long cavity ring up time prevents us from performing pulsed Rabi oscillations, though this could be solved using fast $B_0$ pulses to tune the spin resonance frequency. Our demonstration of $\Delta m_s = \pm 1$ transitions by phonons enables direct PL contrast (optical detection) of resonant spin-strain coupling for sensing applications without electromagnetic microwaves.

To complement $\Delta m_s = \pm 1$ spin driving, we further use the strain coupling terms $S_x^2 - S_y^2$ and $\{S_x, S_y\}$ in the zero-field splitting Hamiltonian to show $\Delta m_s = \pm 2$ spin transitions. For these transitions, PL contrast from ODMR cannot directly measure resonant strain without extra electromagnetically driven spin resonance because PL contrast is insensitive to differences between $|+1\rangle$ and $|-1\rangle$ states. The mechanical transition rate ($\Omega_m$) is instead measured using Autler-Townes (AC Stark) splittings, where in the dressed basis, the new eigenstates are split in energy by $\Omega_m$. This splitting can be observed in the ODMR spectrum. We use a continuous magnetic microwave pump (Rabi frequency $\Omega_{B:\pm 1} \sim$ MHz) for $|0\rangle$ to $|\pm 1\rangle$ transitions while the SAW is driven at a constant frequency $\omega_m/2\pi$ (Fig. 3a). Dressed state level anti-crossings are most clearly seen when the $|\pm 1\rangle$ spin sublevels are tuned to the SAW resonance frequency. The dressed spin eigenstate energies observed for a 400 mW RF drive on the Gaussian SAW resonator closely match predictions[13] for $\Omega_m/2\pi \approx 4$ MHz (Fig. 3b). Additionally, the Autler-Townes splitting scales linearly with square-root of RF power delivered to the SAW, which is expected as $\Omega_m$ is linearly proportional to strain (Fig. 3c). The resolved Autler-Townes splitting shows that the mechanical drive rate is faster than the ensemble spin inhomogeneous linewidth (decoherence rate), allowing for measurement of coherent Rabi oscillations.

We mechanically drive coherent Rabi oscillations of *kk* electron spins using the pulse sequence in Fig. 3d to differentiate between populations transferred to $|+1\rangle$ versus $|-1\rangle$ spin states.

The spin ensemble inhomogeneous linewidth (~1 MHz) and relatively long cavity ring up time ($2Q/\omega_m \approx 16$ μs) prevent fast mechanical pulsing, so we keep the mechanical drive on continuously. A pair of magnetic microwave π pulses defines the effective mechanical pulse time τ seen by the spin ensemble (Fig. 3d). Using this pulse sequence and positive ODMR contrast of *kk* defects, normalized PL values of ±1 can be interpreted as $|\mp 1\rangle$ spin populations, respectively, before the second magnetic π pulse. We find that three-level system dynamics are necessary to explain the observed mechanical Rabi oscillations shown in Fig. 3e, in particular the ensemble population at τ = 0. The observed Rabi oscillations qualitatively agree with spin simulations predicted using only the ensemble ODMR spectrum, fitted Autler-Townes splitting, and numerical modeling for inhomogeneous mechanical driving as a function of depth from Rayleigh waves (Fig. S5). This demonstrates we can mechanically drive $\Delta m_s = \pm 2$ transitions with a Rabi frequency about four times greater than the ensemble ODMR linewidth. Short Rabi decay times are primarily explained by SAW strain inhomogeneity across the ensemble, though another source of damping may be present in the experiments. Coherent Rabi oscillations in ensembles for quantum phononics applications could be improved by using higher quality material and controlled aperture implantations[25] for more homogeneous strain distributions.

We spatially map the Gaussian SAW mode in order to show that $\Delta m_s = \pm 2$ transitions occur due to the mechanical driving and not due to any stray electromagnetic fields[26]. We map changes in the Autler-Townes splitting, shown in Fig. 4a, at a constant magnetic field while sweeping the laser position across the SAW beam waist. In the resonator's transverse direction, a clear Autler-Townes splitting maximum, and therefore resonant strain amplitude, is observed at the Gaussian acoustic focus. $\Omega_m$ as a function of transverse position is well described by a model Gaussian beam waist of the fundamental mode in the device geometry (FWHM = 3.3λ) and not due to predicted stray electric fields (Fig. S4). Scanning the laser spot longitudinally (Fig. 4b), along the SAW propagation, reveals oscillations in the Autler-Townes splitting at the resonator's acoustic half wavelength. Surprisingly, in conflict with assumptions of a simple sinusoidal standing wave containing uniaxial strain nodes (Fig. 1d,e), we observe the mechanical drive rate oscillations are less than 5% peak-to-peak. This is contrary to expectations from previous theoretical work[27] neglecting the full strain tensor, so we interpret our experimental results using a spin Hamiltonian under anisotropic strains also including shear.

The spatial mapping results can be understood by employing a combination of finite-element simulations in conjunction with DFT calculations of spin-strain interactions (see supplementary materials). The $\{\bar{1}\bar{1}20\}$ mirror plane symmetry in 4H-SiC is broken by shears $\varepsilon_{xz}$ and $\varepsilon_{xy}$, which drive the spin out-of-phase with $\varepsilon_{xx} - \varepsilon_{yy}$, $\varepsilon_{yz}$ (mirror symmetry preserving). In our experiments, the Gaussian SAW beamwaist is oriented to propagate in the $\{1\bar{1}00\}$ plane (defined as the xz-plane). The mechanical transition rate is $\Omega_m = \frac{1}{2}(G_{11} - G_{12})\varepsilon_{xx} - 2iG_{14}\varepsilon_{xz}$ corresponding to $\Delta m_s = \pm 2$ transitions, where the spin-strain coupling tensor $G$ is written in Voigt notation. In Fig. 4c we show finite element simulation results for uniaxial strain $\varepsilon_{xx}$ and shear $\varepsilon_{xz}$ for a Rayleigh wave propagating along the x direction. We model the experimental results by converting the strain maps to $\Omega_m$ using $G$ calculated from DFT, which is then convolved with both an optical point-spread function and estimated spatial distribution of the spins (Fig. S5). In our model, spatial averaging causes the spin ensemble to experience similar transition rate $\Omega_m$ magnitudes from $(G_{11} - G_{12})\varepsilon_{xx}$ and $G_{14}\varepsilon_{xz}$ contributions at their respective spatial maxima. These uniaxial strain and shear components, which are spatially offset, do not interfere destructively since $\Omega_m$ is proportional to a linear combination of $\varepsilon_{xx}(S_x^2 - S_y^2)$ and $\varepsilon_{xz}(S_xS_y + S_yS_x)$. Consequently, in qualitative agreement with our calculations (Fig. S6), we always experimentally measure a non-zero Autler-Townes splitting in Fig. 4b. Furthermore, our model explains the relative $\Omega_m$ amplitudes between kk and hh (4.0:1.1) observed in Fig. 4d, and the results for $\Delta m_s = \pm 2$ transitions are well described by the zero-field splitting tensor when the full strain tensor is taken into account. Lastly, we measure mechanical-spin driving on the PL6 defect species in SiC, previously used to demonstrate electron-nuclear spin entanglement in ambient conditions[28]. We find that PL6 experiences similar mechanical transition rates compared to *hh* and *kk* (Fig. 4d); therefore, mechanical control of SiC spin ensembles should be possible at room temperature.

In summary, we have established a Gaussian surface acoustic wave platform for ground state spin control. We imaged the phononic modes using a novel nanoscale X-ray imaging technique and developed a general model of anisotropic lattice perturbations based on *ab initio* calculations, where local defect symmetries are critical to understanding spin-phonon interactions. Surprisingly, shear and uniaxial strain couple to the ground state spin with equivalent magnitudes and different relative phases depending on the strain tensor component. This property could be used to engineer material and device designs that capitalize on mechanical interactions. Since a complete model of spin-strain coupling with $C_{3v}$ symmetry requires six independent coupling parameters, strain cannot necessarily be treated as an equivalent electric field vector. In order to further enhance defect-phonon interaction strengths for hybrid quantum systems, defect excited state electronic orbitals[29,30] and spins[31] could be utilized as opposed to ground state spins[32], and strain effects on defect hyperfine couplings have not been well explored. Our combined theoretical understanding and demonstrations of spin-strain coupling with SiC divacancies provide a basis for quantum sensing with MEMS[33] as well as engineering strong interactions with single phonons for quantum transduction[1], spin squeezing[34], and phonon cooling[35] applications.

## Methods:
### Sample fabrication
The substrate was an on-axis, high purity semi-insulating (HPSI) 4H-SiC commercial wafer (Cree Inc. serial no. W4TRF0R-0200). Defects were created in the 4H-SiC wafer by carbon-12 implantation with dose $10^{12}$ cm$^{-2}$ at 170 keV and 7° tilt (vacancies generated ~300 nm deep from SRIM calculations), however there was a high defect density in the SiC bulk from material growth. After annealing the wafer at 900 °C in $N_2$ for 2 hours, the substrate was then cleaned sequentially with organic solvents, nanostrip, and HF BOE before AlN was sputtered 500 nm thick on the wafer Si face by OEM Group Inc. The AlN layer had a -42 MPa film stress with an AlN [0002] rocking curve 1.52° full width at half maximum (XRD). The interdigitated transducer (IDT), consisting of 80 finger pairs with a window at the SAW focus spanning three wavelengths of missing fingers, was fabricated with 10 nm Ti and 150 nm Al. The Ti/Al and AlN device layers were each dry etched by inductively coupled plasma (ICP) with 10 sccm Ar, 30 sccm $Cl_2$, and 30 sccm $BCl_3$, 400 W ICP power. AlN grooves (650 strips for each grating) in all SAW devices were etched nominally 180 nm deep.

The SAW resonator for spin transitions Figs. 1,3-4 was oriented longitudinally (SAW propagation direction) along $[\bar{1}\bar{1}20]$, and the resonator for Fig. 2 was oriented along $[1\bar{1}00]$. The *x*, *y*, *z* crystal directions are $[\bar{1}\bar{1}20]$, $[1\bar{1}00]$, $[0001]$, respectively. Both devices had Gaussian geometry parameters λ= 12 μm, $w_0$ = 2λ, containing 80 electrode finger pairs and 650 grating strips. All IDT electrodes and grating strips were overlapped 3σ = 3w/√2 in the transverse direction, where w is the Gaussian spot size[36] along the axis of SAW propagation, while electrodes were only apodized by 2σ. Since the X-ray imaging experiments require specific frequencies for stroboscopic X-ray diffraction, results in Fig. 1d,e used a similar IDT geometry, without AlN grooves for increased frequency bandwidth though more impedance mismatched. Fig. 1d,e used a Gaussian IDT with λ = 19.03 μm, $w_0$ = 1.25λ, 120 electrode finger pairs, yielding an IDT center frequency $f_0$ ≈ 352 MHz with 1 MHz bandwidth. This was purposefully matched to an integer multiple of the Advanced Photon Source electron bunch frequency (~ 88 MHz) at Argonne National Laboratory.

## Measurements

All spin manipulation experiments in this study were carried out in a closed cycle cryostat from Montana Instruments Corp. operated with the sample temperature at 30 K. The sample is illuminated using 405 nm and 976 nm laser diodes, with the 405 nm acting as a charge state reset[37] to $VV^0$ and 976nm exciting photoluminescence (> 1000 nm) and initializing the spin state. For pulsed laser experiments, the 976 nm laser is modulated using an acousto-optic modulator (< 50 ns rise time), while the 405 nm laser diode is directly modulated by a current driver (250 kHz bandwidth). Emitted PL was separated by a dichroic and passed through a 1000 nm long pass filter, and then collected into a 62.5 μm core optical fiber. Measurements were realized with an InGaAs photodiode at 1 kHz bandwidth, combined with a lock-in amplifier set at a reference frequency of ~400 Hz for all experiments.

Spin ensembles of $VV^0$ near the Gaussian resonator's beam waist (defined by the acoustic focal spot $w_0$) are initialized and read out optically. To tune the ground state spin sublevels, we use a combination of a permanent magnet and a wire loop on a printed circuit board to produce static magnetic fields ($B_0$) oriented along the 4H-SiC c-axis. A coplanar waveguide (CPW) behind the sample provides microwaves for in-plane magnetic spin control.

## Acknowledgements
The devices and experiments were supported by the Air Force Office of Scientific Research, material for this work was supported by the Department of Energy (DOE). SXDM measurements were performed at the Hard X-ray Nanprobe Beamline, operated by the Center for Nanoscale Materials at the Advanced Photon Source, Argonne National Laboratory (Contract No. DE-AC02-06CH11357).

S.J.W. and K.J.S. were supported by the NSF GRFP, C.P.A. was supported by the Department of Defense through the NDSEG Program, and M.V.H., F.J.H., A.N.C., G.G., D.D.A. were supported by the DOE, Office of Basic Energy Sciences. This work made use of the UChicago MRSEC (NSF DMR-1420709) and Pritzker Nanofabrication Facility, which receives support from the SHyNE, a node of the NSF's National Nanotechnology Coordinated Infrastructure (NSF ECCS-1542205). The authors thank P. J. Duda, P. V. Klimov, P. L. Yu, S. A. Bhave, H. Seo, and N. Schine for insightful discussions and B. B. Zhou, S. Bayliss, and A. L. Yeats for careful reading of the manuscript.

## Author contributions

S.J.W. fabricated devices. S.J.W. and G.W. performed the experiments and data analysis. C.P.A. and A.B. processed materials. H.M. and M.Y. performed density functional theory calculations. K.J.S. and G.K. helped with device characterization. M.V.H. executed X-ray imaging experiments. F.J.H., A.N.C., D.I.S., G.G., and D.D.A. advised on all efforts. All authors contributed to discussions and production of the manuscript.

## Data and materials availability

All data are available upon request to the corresponding author.

## Additional information

The authors declare no competing interests.

Figures

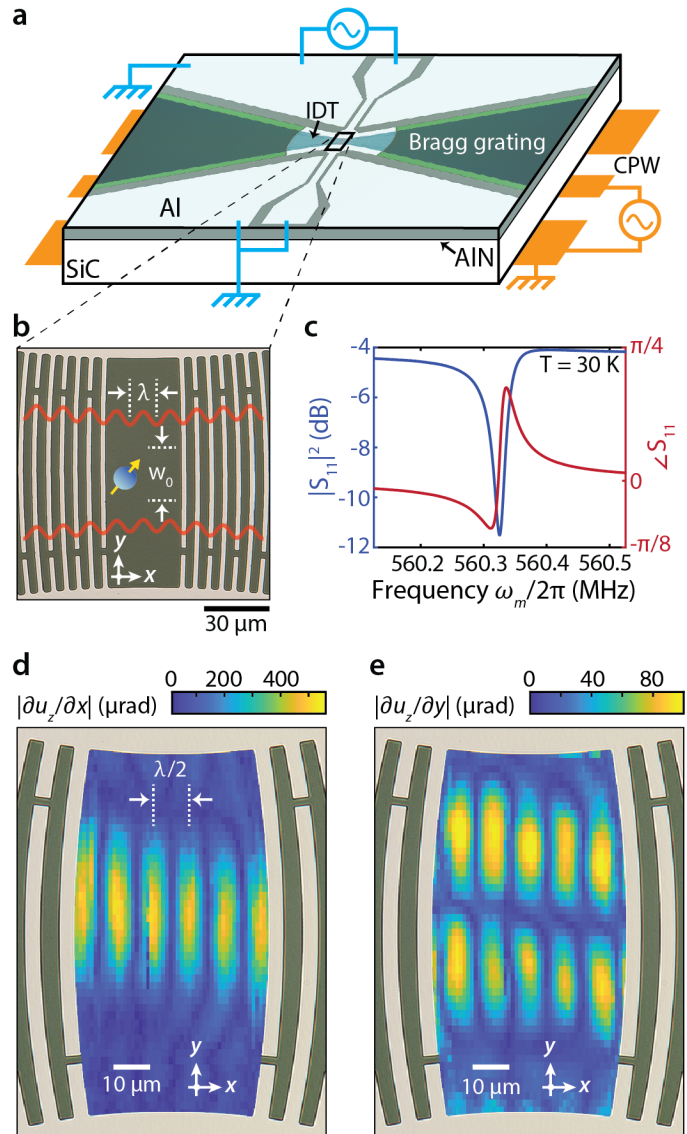

**Fig. 1. Strain focusing with a Gaussian SAW resonator.** (**a**) Schematic of the SAW device geometry fabricated on sputtered AlN on a 4H-SiC substrate. Microwaves drive spin transitions mechanically through the SAW resonator (cyan) and magnetically from the backside coplanar waveguide (orange). (**b**) Optical micrograph of the Gaussian SAW resonator's acoustic focus ($\lambda$ = 12 µm, $w_0$ = 2$\lambda$) with red lines illustrating the wave's out-of-plane displacement ($u_z$). (**c**) Magnitude (blue) and phase (red) measurements of the 1-port RF reflection of the Gaussian SAW resonator used in spin experiments. (**d,e**) Mechanical mode from a similar Gaussian SAW ($\lambda$ = 19 µm, $w_0$ = 1.25$\lambda$), directly measured with s-SXDM using the 4H-SiC [0004] Bragg peak. This quantifies the SAW peak-to-peak (d) longitudinal and (e) transverse lattice slopes at the acoustic beam waist. The image is skewed vertically due to sample drift during measurements.

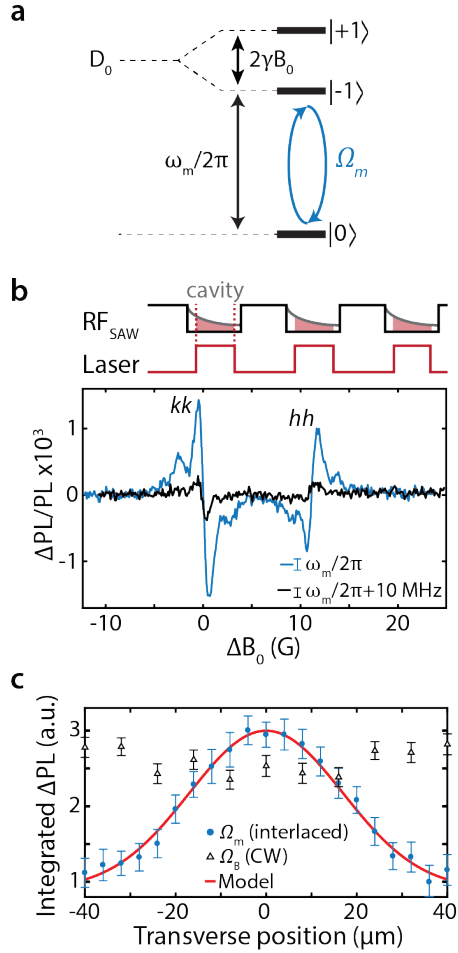

Fig. 2. **Optically detected acoustic paramagnetic resonance in silicon carbide.** (**a**) Energy level diagram showing the SAW frequency on resonance with the spin transition between the $|0\rangle$ and $|-1\rangle$ states. (**b**) (Top) Interlaced pump-probe sequence during magnetic field-modulation. (Bottom) PL contrast at 30 K when RF excitation is on and off cavity resonance ($\omega_m/2\pi$ = 559.6 MHz). RF power is 32 mW at sample, and $\Delta B_0$ is in reference to the drive frequency. (**c**) Integrated PL contrast from *kk* resonance (evaluated at $\Delta B_0 = 0$) as a function of the SAW resonator transverse position. RF on-resonance ("$\Omega_m$") uses the interlaced sequence from (b), whereas off-resonance data ("$\Omega_B$") uses a continuous, non-interlaced sequence. RF power is 200 mW at sample, and the beam waist model is $\exp(\frac{-y^2}{w_0^2})$, using fabrication parameters and a scaled amplitude. All error bars are 95% confidence intervals.

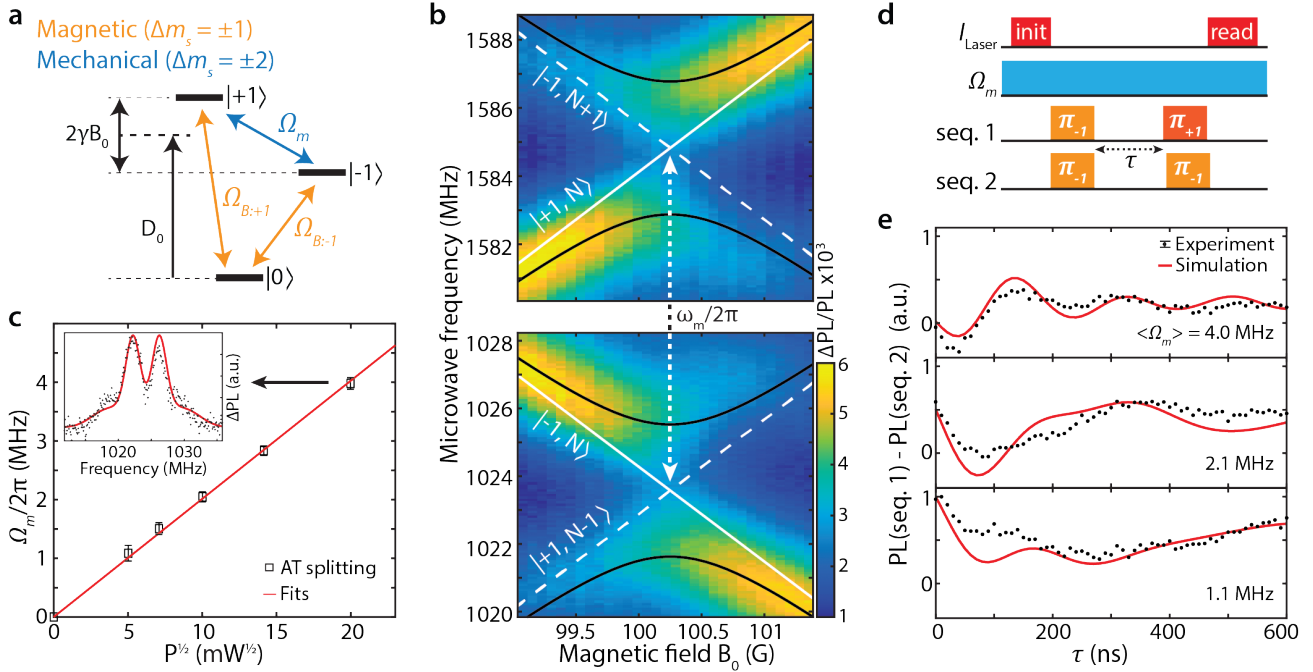

**Fig. 3. Coherent mechanical driving of *kk* spin ensembles.** (**a**) VV ground state illustration with magnetic ($\Omega_{B:\pm 1}$) and electromechanical ($\Omega_m$) drives shown. (**b**) Autler-Townes measurement on a *kk* ensemble at 30 K; dressed for *N* phonons (black) and undressed (white) spin transitions. The mechanically dressed eigenstates and corresponding transitions are split by $\Omega_m$. (**c**) Mechanical transition rates obtained from Autler-Townes splittings agree with a linear fit to the square-root of drive power. Error bars are 95% confidence intervals from fits. Inset shows an Autler-Townes splitting measurement (black) at $B_0 \approx 100$ G, with Gaussian fits (red) to the $VV^0$ electron spin and weakly coupled nearby nuclear spins. (**d**) Pulse sequence for mechanically driven Rabi oscillations. (**e**) Mechanically driven Rabi oscillations at ~ 400, 100, and 25 mW, respectively, and typical error bars are 95% confidence intervals. The signal for each Rabi oscillation is normalized by a global factor, and simulations are ensemble average predictions with inhomogeneous strain distributions from finite element modeling.

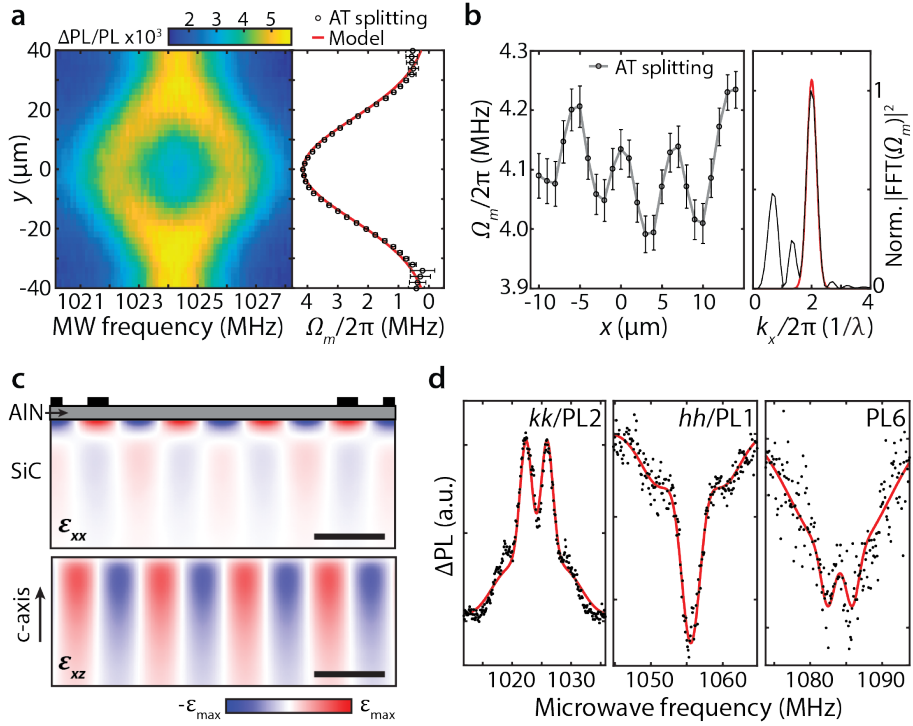

Fig. 4. Spatially mapping mechanical spin drive rates and defect comparisons. (a) Autler-Townes splitting (left) of $kk$ $|-1\rangle$ sublevel as a function of transverse position at $x = 0$ and mechanical transition rate (right) analyzed from Autler-Townes splittings. The beam waist model only uses fabrication parameters with a scaled amplitude. (b) Mechanical transition rate (left) as a function of longitudinal position at $y = 0$, plotted with a line through the experimental data. FFT (right) shows a peak and Gaussian fit in red at the expected acoustic periodicity $\lambda/2$ (6 μm). (c) Strain $\varepsilon_{xx}$ and $\varepsilon_{xz}$ of the SAW modeled with COMSOL Multiphysics. (d) Autler-Townes splitting measurements for $kk$, $hh$, and PL6 with $\Omega_m \sim$ 4.0, 1.1, 3.4 MHz, respectively, under the same conditions. All error bars are 95% confidence intervals from fitting and measurements are performed at 30 K.